\documentclass{procmult}
\usepackage{amsmath}
\usepackage{amsfonts}
\usepackage[dvips]{graphicx}

\begin{document}

\newcommand\be{\begin{equation}}
\newcommand\ee{\end{equation}}
\newcommand\bea{\begin{eqnarray}}
\newcommand\eea{\end{eqnarray}}
\newcommand\bseq{\begin{subequations}} 
\newcommand\eseq{\end{subequations}}
\newcommand\bcas{\begin{cases}}
\newcommand\ecas{\end{cases}}
\newcommand{\p}{\partial}
\newcommand{\f}{\frac}

\title{GUP vs polymer quantum cosmology: the Taub model}
\author{Marco Valerio Battisti$^\dag$, Orchidea Maria Lecian$^\ddag$ \and Giovanni Montani$^\S$}
\institute{\small{
$^\dag$ Centre de Physique Th\'eorique de Luminy, Universit\'e de la M\'editerran\'ee, F-13288 Marseille FR, and ``Sapienza'' Universit\`a di Roma, Dipartimento di Fisica and ICRA,\\ P.le A. Moro 5, 00185 Rome IT, battisti@icra.it \\
$^\ddag$ Institut des Hautes Etudes Scientifiques, 91440 Bures-sur-Yvette FR, and ``Sapienza'' Universit\`a di Roma, Dipartimento di Fisica and ICRA, P.le A. Moro 5, 00185 Rome IT, lecian@ihes.fr \\
$^\S$ ICRA, ICRANet, ENEA and ``Sapienza'' Universit\`a di Roma, Dipartimento di Fisica, P.le A. Moro 5, 00185 Rome IT, montani@icra.it} 
}

\maketitle

\abstract{The fate of the cosmological singularity in the Taub model is discussed within the two frameworks. An internal time variable is ruled out and the only remaining degree of freedom (the anisotropy) of the Universe is quantized according to such schemes. The resulting GUP Taub Universe is singularity-free, differently from the second case, where the classical singularity is not tamed by the polymer-loop quantum effects.}

\bigskip

\section{Introduction} 

Two different quantum cosmology approaches are applied to the Taub model. The study is performed at the classical and at the quantum level in both schemes. In particular, the generalized uncertainty principle (GUP) and the polymer (loop) frameworks are implemented to this system. In the first case \cite{BM08}, the cosmological singularity appears to be probabilistically suppressed, while, in the second one \cite{BLM08}, the Universe is still singular. Such a feature then allows us to better understand the avoidance of the cosmological singularity in other different quantum gravity toy models.

The Taub Universe arises as a particular case of the Bianchi IX model, i.e. the most general scheme allowed by the homogeneity constraint (for reviews see \cite{rev}). It is obtained by restricting the dynamics to that of a one-dimensional particle bouncing against a wall, when only one degree of freedom (the Universe anisotropy) is taken into account. The relevance of the Taub Universe in quantum cosmology is then due to the fact that it is a necessary step towards the Bianchi IX model, being a generalization of other isotropic models. In particular, it has been used to test the validity of the minisuperspace scheme \cite{kuch} and to explore the application of the extrinsic cosmological time \cite{altri}. 

The paper is organized as follows. In Section 2 the Taub model and the formalisms are reviewed. Section 3 and 4 are devoted to the classical and quantum analysis respectively. Comparisons with other approaches follow in Section 5. We adopt $\hbar=c=16\pi G=1$ units.

\section{Taub Universe and formalisms}

The Taub cosmological model is a particular case of the Bianchi IX Universe, of which only one anisotropy is taken into account. The Bianchi IX model, together with Bianchi VIII, is the most general spatially-homogeneous model and is described by the line element $ds^2=N^2dt^2-e^{2\alpha}\left(e^{2\gamma}\right)_{ij}\omega^i\otimes\omega^j$ \cite{rev}. Here $N=N(t)$ is the lapse function, $\omega^i=\omega^i_adx^a$ are the $SO(3)$ left-invariant 1-forms, $\alpha=\alpha(t)$ describes the isotropic expansion of the Universe and $\gamma_{ij}=\gamma_{ij}(t)$ is a traceless symmetric matrix, which determines the anisotropies via $\gamma_\pm$. The dynamics of this Universe towards the singularity is described by the motion of a two-dimensional particle (the two physical degree of freedom of the gravitational field) in a dynamically-closed domain \cite{rev}. In the Misner picture, such a domain depends on the time variable $\alpha$, while, in the Misner-Chitr\'e one, it becomes stationary in time. Performing the ADM reduction of the dynamics (according to which the classical constraints are solved with respect to the given momenta before implementing any quantization algorithm), an effective Hamiltonian is obtained, which depends only on the physical degrees of freedom of the system. In particular, the scalar constraint is solved with respect to the momentum conjugated to the time variable $p_\tau$ (we adopt the time gauge $\dot\tau=1$) and, performing another change of variables, we obtain
\be\label{huv}
-p_\tau\equiv H^{IX}_{ADM}=v\sqrt{p_u^2+p_v^2}.
\ee 
The dynamics of such a system is equivalent to a billiard ball on a Lobatchevsky plane and the three corners of the Misner scheme are replaced by the points $(0,0)$, $(-1,0)$ and $v\rightarrow\infty$ in the $(u,v)$-plane, as in Fig. 1.
\begin{figure}
\centering
\includegraphics[height=1.5in]{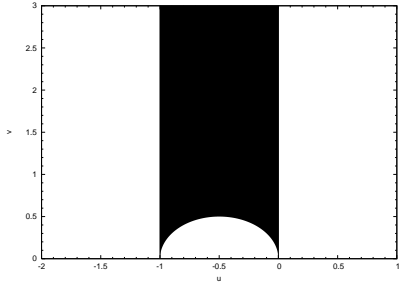}
\caption{\small{The dynamical-allowed domain in the $(u,v)$-plane where the dynamics is restricted.}} 
\end{figure} 
The Taub cosmological model is described by $\gamma_-=0$. The dynamics of this Universe is equivalent to the motion of a particle in a one-dimensional closed domain. Such a domain corresponds the choice of only one of the three equivalent potential walls of the Bianchi IX model. The ADM Hamiltonian (\ref{huv}) rewrites $H_{ADM}^T=vp_v$, where $v\in[1/2,\infty)$. This Hamiltonian can be further simplified defining the new variable $x=\ln v$, and becomes
\be\label{ht}
H_{ADM}^T=p_x\equiv p
\ee
where the configuration variable $x$ is related to the Universe anisotropy $\gamma_+$ by the equation $\gamma_+=e^{\tau-x}(e^{2x}-3/4)/\sqrt3$. The Hamiltonian (\ref{ht}) will be the starting point of our analysis. It is worth noting that the classical singularity now appears for $\tau\rightarrow\infty$. 

\subsection{GUP quantum mechanics}

Some issues and results of a non-relativistic quantum mechanics with non-zero minimal uncertainties in position are briefly reviewed \cite{Kem}. In one dimension, we consider the Heisenberg algebra generated by $q$ and $p$ obeying the commutation relation  
\be\label{modal}
[q,p]=i(1+\beta p^2), 
\ee
where $\beta>0$ is a deformation parameter. This commutation relation leads to the generalized uncertainty relation $\Delta q \Delta p\geq \f 1 2\left(1+\beta (\Delta p)^2+\beta \langle p\rangle^2\right)$, which appears in string theory \cite{String}. The canonical Heisenberg algebra can be recovered in the limit $\beta=0$, and the generalization to more dimension is straightforward, leading naturally to a ``noncommutative geometry'' for the space coordinates.

It is immediate to verify that a finite minimal uncertainty in position $\Delta q_{min}=\sqrt\beta$ is predicted. The existence of a non-zero uncertainty in position is relevant since it implies that there cannot be any physical state that is a position eigenstate. In fact, an eigenstate of an observable necessarily has a vanishing uncertainty on it. Although it is possible to construct position eigenvectors, they are only formal eigenvectors and not physical states. The deformed Heisenberg algebra (\ref{modal}) can be represented in the momentum space as 
\be\label{rep}
p\psi(p)=p\psi(p), \qquad q\psi(p)=i(1+\beta p^2)\p_p\psi(p),
\ee
on a dense domain $S$ of smooth functions. To recover information on position we have to study those states, which realize the maximally-allowed localization. Such states $\vert\psi^{ml}_{\zeta}\rangle$, which are proper physical states around a position $\zeta$, have the proprieties $\langle\psi^{ml}_{\zeta}\vert q\vert\psi^{ml}_{\zeta}\rangle=\zeta$ and $(\Delta q)_{\vert\psi^{ml}_{\zeta}\rangle}=\Delta q_{min}$. We can project an arbitrary state $\vert\psi\rangle$ on the maximally-localized states $\vert\psi^{ml}_{\zeta}\rangle$ to obtain the probability amplitude for a particle being maximally localized around the position $\zeta$ (i.e. with standard deviation $\Delta q_{min}$). We call these projections the {\it quasiposition wave function} $\psi(\zeta)=\langle\psi^{ml}_{\zeta}\vert\psi\rangle$, and explicitly we have the generalized Fouries transformation
\be\label{qwf} 
\psi(\zeta)\sim\int^{+\infty}_{-\infty}\f{dp}{(1+\beta p^2)^{3/2}} \exp\left(i\f{\zeta}{\sqrt{\beta}} \tan^{-1}(\sqrt{\beta}p)\right)\psi(p).
\ee
As $\beta\rightarrow0$, the ordinary position wave function $\psi(\zeta)=\langle\zeta\vert\psi\rangle$ is recovered.

\subsection{Polymer quantum mechanics}

The polymer representation of quantum mechanics consists in defining abstract kets, labeled by a real number and assumed to form an orthonormal basis, and then considering a suitable finite subset of them, whose Hilbert space is defined by the corresponding inner product \cite{pol}. This procedure helps one gain insight onto some particular features of quantum mechanics, when an underlying discrete structure is somehow hypothesized. The request that the Hamiltonian associated to the system be of direct physical interpretation defines the polymer phase space, and the continuum limit can be recovered by the introduction of the concept of the scale \cite{cori2}.

In the particular case of a discrete position variable in the momentum polarization, the Hamiltonian variable $p$ cannot be implemented as an operator, so that some restrictions on the model have to be required. If the set of kets is restricted by the introduction of a regular graph $\gamma_{\mu_0}$, the kynetic term of the Hamiltonian is approximated by the polymer substitution
\begin{equation}\label{para}
p\rightarrow \frac{1}{\mu_0}\sin (\mu_0p),
\end{equation}
where the incremental ratio is evluated for an exponentiated operator. The Hamiltonian operator $H_{\mu_0}$, which lives in $\mathcal{H}_{\gamma_{\mu_0}}$, reads 
\begin{equation}
H_{\mu_0}=\frac{\hat{p}_{\mu_0}^2}{2m}+V(\hat{q}).
\end{equation}
The definition of a scale, $C_n$, eables one to approximate continuous functions with functions that are constant on the intervals. As a result, at any given scale $C_n$, the kinetic term of the Hamiltonian operator can be approximated, and effective theories at given scales are related by coarse-graining maps.

\section{Deformed classical dynamics}

The ordinary Taub model can be interpreted as a massless scalar relativisitic particle moving in the Lorentzian minisuperspace ($\tau,x$)-plane, whose the classical trajectory is its light-cone. More precisely, the incoming particle ($\tau<0$) bounces on the wall ($x=x_0=\ln(1/2)$) and falls into the classical cosmological singularity ($\tau\rightarrow\infty$). Investigating the modification of the dynamics within the two frameworks will show that the two behaviors can be interpreted as complementary.

\subsection{GUP framework}

The GUP-classical dynamics is contained in the modified symplectic geometry arising from the classical limit of (\ref{modal}), as soon as the parameter $\beta$ is regarded as an independent constant with respect $\hbar$. It is then possible to replace the quantum-mechanical commutator (\ref{modal}) via its Poisson brackets, i.e. $-i[q,p]\Longrightarrow\{q,p\}=(1+\beta p^2)$. The Poisson brackets for any two-dimensional phase space function are
\be
\{F,G\}=\left(\f{\p F}{\p q}\f{\p G}{\p p}-\f{\p F}{\p p}\f{\p G}{\p q}\right)(1+\beta p^2).
\ee
Applying this scheme to the Hamiltonian (\ref{ht}), we immediately obtain the equations of motion for the model \cite{BM08},
\be\label{moteq}
x(\tau)=(1+\beta A^2)\tau+cost, \qquad p(\tau)=cost=A,
\ee
where $x\in[x_0,\infty)$. Therefore, at the classical level, the effects of the deformed Heisenberg algebra (\ref{modal}) on the Taub Universe are as follows. The angular coefficient is $(1+\beta A^2)>1$ for $\beta\neq0$, and thus the angle between the two straight lines $x(\tau)$, for $\tau<0$ and $\tau>0$, becomes smaller as the values of $\beta$ grows. The trajectories of the particle (Universe), before and after the bounce on the potential wall at $x=x_0\equiv\ln(1/2)$, are closer to each other then in the canonical case ($\beta=0$). 

\subsection{Polymer framework}

The polymer-classical dynamics relies on the substitution (\ref{para}) in the Hamiltonian of the model (\ref{ht}). This way, the equations of motion rewrite \cite{BLM08}
\be
\dot{x}=\left\{x,H\right\}=\cos(ap), \qquad \dot{p}=\left\{p,H\right\}=0,
\ee
where a dot denotes differentiation with respect to the time variable $\tau$. The equations of motion are immediately solved as
\be\label{eqmot}
x(\tau)=\cos(ap) \tau, \qquad p(\tau)=A,
\ee
where $A$ is a constant. In the discretized (polymer) case, i.e. for $a\neq0$, the one-parameter family of trajectories flattens. In fact, the angle between the incoming trajectory and the outgoing one is greater than $\pi/2$, since $p\in\left(-\pi/a, \pi/a\right)$. As these trajectories diverge rather than converge, we expect the polymer quantum effects to be reduced with respect to the classical case, as we will verify below. 

\section{Deformed quantum dynamics}

The quantum dynamics of the Taub Universe is here investigated according to the two different approaches. Particular attention is paid to the wave-packet evolution and the consequential fate of the classical cosmological singularity. In both frameworks, the variable $\tau$ is regarded as a time coordinate and therefore ($\tau,p_\tau$) are treated in the canonical way. The deformed quantization (GUP or polymer) is then implemented only to the submanifold describing the only degree of freedom of the Universe, i.e. the phase space spanned by ($x,p$). We then deal with a Schr\"odinger-like equation
\be
i\p_\tau\Psi(\tau,p)=\hat H_{ADM}^T\Psi(\tau,p),
\ee
where the operator $\hat H_{ADM}^T$ accounts for the modifications due to the two frameworks. We have to square the eigenvalue problem in order to correctly impose the boundary condition: in agreement with the analysis developed in \cite{Puzio}, we make the well-grounded hypothesis that the eigenfunctions form be independent of the presence of the square root, since its removal implies the square of the eigenvalues only. The wave packets, which are superposition of the eigenfunctions $\Psi(\tau,x)=\int_0^\infty dk A(k)\psi_k(x)e^{-ik\tau}$, are then constructed for both models, taking $A(k)$ as a Gaussian-like weighting function. The differences between the two approaches are due to the features of the eigenfunctions $\psi_k(x)$. Analyzing such an evolution, we show that the GUP Taub Universe appears to be probabilistically singularity-free, differently from the polymer case, where the singularity is not tamed by the cut-off-scale effects.

\subsection{GUP framework}

We now analyze the model in the GUP approach \cite{BM08}. As explained before, we lost all informations on the position itself, so that the boundary conditions have to be imposed on the quasiposition wave function (\ref{qwf}), i.e. $\psi(\zeta_0)=0$ (where $\zeta_0=\langle\psi^{ml}_{\zeta}\vert x_0\vert\psi^{ml}_{\zeta}\rangle$, in agreement with the previous discussion). The solution of the eigenvalue problem is the Dirac $\delta$-distribution $\psi_k(p)=\delta(p^2-k^2)$, and therefore the quasiposition wave function (\ref{qwf}) reads 
\be\label{ef}
\psi_k(\zeta)=\f A{k(1+\beta k^2)^{3/2}}\left[\exp\left(i\f{\zeta}{\sqrt{\beta}} \tan^{-1}(\sqrt{\beta}k)\right)-\exp\left(i\f{(2\zeta_0-\zeta)}{\sqrt{\beta}} \tan^{-1}(\sqrt{\beta}k)\right)\right],
\ee
where $A$ is a constant and the boundary condition $\psi(\zeta_0)=0$ has been imposed. The deformation parameter $\beta$, i.e. the presence of a non-zero minimal uncertainty for the configuration variable, is responsible for the GUP effects on the dynamics. The physical interpretation of $\beta$ is then a non-zero minimal uncertainty in the anisotropy of the Universe. To better understand the modifications induced by the deformed Heisenberg algebra on the canonical Universe dynamics, we have to analyze different $\beta$-regions. In fact, when $\beta$ becomes more and more important, i.e. when we are at some scale that allows us to appreciate the GUP effects, the evolution of the wave packets is different from the canonical case. More precisely, these effects are present when the product $k_0\sqrt\beta$ becomes remarkable, i.e. when $k_0\sqrt\beta\sim\mathcal O(1)$, and therefore when $\beta$ is comparable to $1/k_0^2$. In fact, the correct semiclassical behaviors of the model far away from the singularity is described by wave packets peaked at energies much smaller then $1/\sqrt\beta$ \cite{BM08}. In particular, for $k_0=1$, we can distinguish between three different $\beta$-regimes:
\begin{itemize}
	\item $\beta\sim\mathcal O(10^{-2})$ regime. The wave packets begin to spread and a constructive and destructive interference between the incoming and outgoing wave appears. The probability amplitude to find the Universe is still peaked around the classical trajectory.
	\item $\beta\sim\mathcal O(10^{-1})$ regime. It is no more possible to distinguish an incoming or outgoing wave packet and, at this level, the notion of a wave packet following a classical trajectory becomes meaningless. 
	\item $\beta\sim\mathcal O(1)$ regime. A dominant probability peak ``near'' the potential wall appears. There are also other small peaks for growing values of $\zeta$, but they are widely suppressed for bigger $\beta$. The motion of wave packets shows a stationary behavior, i.e. these are independent of $\tau$. See Fig. 2.
\end{itemize}
Following this picture we are able to learn the GUP modifications to the WDW wave packets evolution. In fact, from small to big values of $\beta$, we can see how the wave packets {\it escape} from the classical trajectories and approach a stationary state close to the potential wall. 
\begin{figure}
\centering
\includegraphics[height=1.5in]{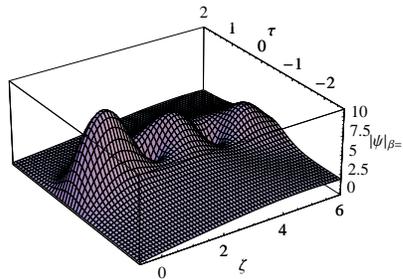}
\caption{Wave packets $\vert\Psi(\tau,\zeta)\vert$ in the GUP framework as $\beta k_0^2=1$ ($k_0=1$ and $\sigma=4$).} 
\end{figure}

Such a behavior is, in some sense, expected from a classical point of view. In fact, at classical level the ingoing and the outgoing trajectories shrink each other. So a quantum probability interference is a fortiori predicted. On the other hand, the stationarity feature exhibited by the Universe in the ($\beta\sim\mathcal O(1)$)-region is a purely quantum GUP effect. Such a behavior cannot be inferred from a deformed classical analysis. From this point of view, the classical singularity ($\tau\rightarrow\infty$) is widely probabilistically suppressed, because the probability to find the Universe is peaked just around the potential wall. This way we claim that the GUP-Taub Universe is singularity-free.

\subsection{Polymer framework}

We now analyze the model in the polymer approach \cite{BLM08}. For the quantum analysis of the model, we choose a discretized $x$ space, and solve the corresponding eigenvalue problem in the $p$ polarization. Considering the time evolution for the wave function $\Psi$ as given by $\Psi_k(p,\tau)=e^{-ik\tau}\psi_k(p)$ and the results of \cite{Puzio}, we obtain the following eigenvalue problem
\be
(p^2-k^2)\psi_k(p)=\left[\frac{2}{a^2}\left(1-\cos(ap)\right)-k^2\right]\psi_k(p),
\ee 
solved by
\begin{subequations}
\begin{align}
&k^2=k^2(a)=\frac{2}{a^2}\left(1-\cos(ap)\right)\leq k^2_{max}=\frac{4}{a^2}\label{kappa}\\
&\psi_{k,a}(p)=A\delta(p-p_{k,a})+B\delta(p+p_{k,a})\label{pi}\\
&\psi_{k,a}(x)=A\left[\exp(ip_{k,a} x)-\exp(ip_{k,a}(2x_0-x))\right]\label{ics}:
\end{align}
\end{subequations}
(\ref{pi}) is the momentum wave function, with $A$ and $B$ two arbitrary integration constant, and (\ref{ics}) is the coordinate wave function, where an integration constant has been eliminated by imposing suitable boundary conditions. Moreover, we have defined the modified dispersion relation
\be\label{disprel}
p_{k,a}\equiv\frac{1}{a}\arccos\left(1-\frac{k^2a^2}{2}\right)
\ee 
from (\ref{kappa}). Furthermore, we stress that $k^2$ is bounded from above, as illustrated in (\ref{kappa}), but it is its square root, considered for its positive determination, which accounts for the time evolution of the wave function.

We now construct suitable wave packets $\Psi(x,\tau)$ taking into account the previous discussion (note that a maximum energy $k_{max}$ is now predicted). Three relevant cases can be distinguished:
\begin{figure}
\centering
\includegraphics[height=1.5in]{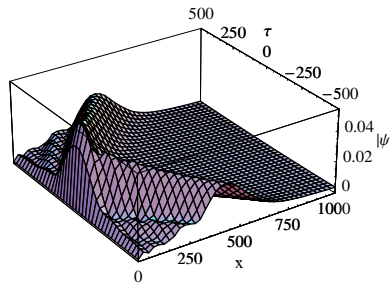}
\caption{The spread polymer wave packet $|\Psi(x,\tau)|$ as $k_0 a=1/2$ ($a=50$, $k_0=0.01$, $\sigma=0.125$).} 
\end{figure}
\begin{itemize}
	\item $k_0a\sim\mathcal{O}(1)$ and peaked weighting function. The resulting wave packet is well approximated by a purely monochromatic wave. A small interference phenomenon between the wave and the wall is then predicted.
	\item $k_0a\sim\mathcal{O}(1)$ and spread weighting function. A strong interference phenomenon between the wave and the wall now appears. Nevertheless, this interference phenomenon is not able to probabilistically tame the singularity, as it takes place in the 'outer' region, in a way complementary to that of the GUP approach (see Fig. 3). The polymer-Taub Universe is then still a singular cosmological model. 
	\item $k_0a\ll\mathcal O(1)$ regime. This can be considered as the semi-classical limit of the model. In fact, differently from the other cases, the value of $k_0$ around which the wave packet is peaked is not arbitrary, but constrained by the characteristic scale $a$ under investigation. The ordinary WDW behavior is therefore recasted.
\end{itemize}

\section{Comparison with other approaches}

The Taub cosmological model offers a suitable scenario, where different quantization techniques can be applied. In fact, it is possible to single out a time variable, so that the anisotropy describes the real degree of freedom of the Universe. It is therefore reasonable to investigate the fate of the cosmological singularity without modifying the time variable. The comparison with analysis of the cosmological singularity in other cosmological models outlines how the features of the Taub model allow one to pick the cut off effects out of those due to the choice of the Hamiltonian variables.
   
In the cosmological isotropic sector of GR, i.e. the FRW models, the singularity is removed by loop quantum effects. The wave function of the universe exhibits a non-singular behavior at the classical singularity, and the big-bang is replaced by a big-bounce, when a free scalar field is taken as the relational time \cite{APS}. The scale factor of the universe is directly quantized by the use of the polymer (loop) techniques, so that the evolution itself of the wave packet of the universe is deeply modified by such an approach. The Hamiltonian constraint does not allow for a constant solution of the variable conjugated to the scale factor, so that it is not possible to choose a scale, such that the polymer modifications are negligible throughout the whole evolution, so that the comparison with the ordinary representation is not always possible. Anyhow, we stress that, for the Taub model, the cosmological singularity is probabilistically suppressed, regardless to the fact whether the system can appreciated or not the cut off during the whole evolution.

In \cite{homo1}, all the degrees of freedom of the Bianchi cosmological models in the ADM reduction of the dynamics are quantized by loop techniques. In particular, also the time variable, i.e. the Universe volume, is treated at the same level as the others. In most cases, the time variable is defined by a phase space variable, i.e. it is an internal one. As a result, also the Bianchi Universes are singularity-free \cite{homo2}. In this respect, our analysis is based on considering the time variable as an ordinary Heisenberg variable, while the cut off is imposed on the anisotropy only. 

The GUP dynamics of other cosmological models has been investigated in different approaches. In particular, the big-bang singularity appears to be tamed by GUP effects showing a stationary behavior of the wave packets \cite{BM07}. Such a prediction is in agreement with those achieved in a noncommutative quantum cosmology \cite{VS}. However, in order to predict a big bounce \`a la LQC, a Snyder-deformed quantum cosmology has to be addressed \cite{B08}. As the last point, it is interesting to notice that the GUP-Mixmaster Universe is still a chaotic model \cite{BM08b}, as opposite to the LQC one \cite{Mixl}, the difference being essentially based on the application of the deformed scheme to the time variable too.      
     
\bigskip     
     
{\it Acknowledgments.} The ``Angelo Della Riccia'' Fellowship and the ``Sapienza CUN 2'' Fellowship are gratefully acknowledged.

\end{document}